\documentclass[journal]{IEEEtran}

\usepackage{cite}
\DeclareFontFamily{OT1}{pzc}{}
\DeclareFontShape{OT1}{pzc}{m}{it}{<-> s * [1.10] pzcmi7t}{}
\DeclareMathAlphabet{\mathpzc}{OT1}{pzc}{m}{it}

\usepackage{xcolor}
\usepackage{multirow}
\usepackage{graphicx,epsfig}
\usepackage{graphics}

\usepackage{flushend}

\usepackage{color, colortbl}
\definecolor{Gray}{gray}{0.9}
\definecolor{LightCyan}{rgb}{0.88,1,1}
\usepackage[first=0,last=9]{lcg}

\usepackage{color, colortbl}
\definecolor{Gray}{gray}{0.9}
\definecolor{LightCyan}{rgb}{0.88,1,1}
\usepackage[first=0,last=9]{lcg}

\usepackage{color}

\ifCLASSINFOpdf
\else
\fi

\usepackage{amsmath}
\usepackage{textcomp}
\usepackage{mathtools}
\usepackage{enumerate}

\usepackage{color}

\usepackage{xcolor}
\usepackage{multirow}
\usepackage{amsbsy,amsfonts,amsfonts,amssymb,epsfig}
\usepackage{graphicx}
\usepackage{graphics}

\makeatletter
\def\vhrulefill#1{\leavevmode\leaders\hrule\@height#1\hfill \kern\z@}
\makeatother

\makeatletter 
\let\myorg@bibitem\bibitem
\def\bibitem#1#2\par{%
	\@ifundefined{bibitem@#1}{%
		\myorg@bibitem{#1}#2\par
	}{%
		\begingroup
		\color{\csname bibitem@#1\endcsname}%
		\myorg@bibitem{#1}#2\par
		\endgroup
	}%
}
\makeatother 

\begin{document}

\title{Antenna Selection For Receive Spatial Modulation System Empowered By Reconfigurable Intelligent Surface}
%
%


\author{Burak Ahmet Ozden, Erdogan Aydin
	
	\thanks{B. A. Ozden is with Istanbul Medeniyet University, Department of Electrical and Electronics Engineering, 34857, Uskudar, Istanbul, Turkey, and also with Y{\i}ld{\i}z Technical University, Department of Computer Engineering, 34220, Davutpasa, Istanbul, Turkey (e-mail: bozden@yildiz.edu.tr)}
	
	\thanks{E. Aydin is with Istanbul Medeniyet University, Department of Electrical and Electronics Engineering (e-mail: erdogan.aydin@medeniyet.edu.tr) (Corresponding author: Erdogan Aydin.)}
	
}

\maketitle

\begin{abstract}
Reconfigurable intelligent surface (RIS) enhances signal quality by adjusting the phase of electromagnetic waves in wireless communication. Spatial modulation (SM), a prominent index modulation (IM) technique, provides high spectral efficiency and low energy consumption. In this article, a new wireless communication system is proposed by combining capacity-optimized antenna selection (COAS), antenna correlation antenna selection (ACAS), and Euclidean distance-optimized antenna selection (EDAS)-supported RIS-empowered receive SM (RIS-RSM) system (AS-RIS-RSM) in a single-input multiple-output (SIMO) structure. The proposed AS-RIS-RSM schemes (COAS-RIS-RSM, ACAS-RIS-RSM, and EDAS-RIS-RSM) have superior features such as high spectral efficiency, high energy efficiency, and low error data transmission. Integrating  COAS, ACAS, and EDAS techniques into the system enables the selection of the channel with the best conditions, thus increasing the error performance of the proposed system. Also, using RIS increases the error performance of the system by controlling the transmitted signal to a certain extent.  The analytical ABER results of the proposed AS-RIS-RSM systems are derived and shown to overlap with simulation results. For the proposed systems, an optimal maximum likelihood (ML) detector and a sub-optimal low-complexity greedy detector (GD) are offered. Also, capacity analyses of the proposed AS-RIS-RSM systems are derived and it is observed that they have higher capacity compared to RIS-QAM/PSK and RIS-RSM systems. Then, computational complexity analyses of the proposed COAS-RIS-RSM, ACAS-RIS-RSM, and EDAS-RIS-RSM systems are presented. The proposed systems have been compared to counterpart wireless communication systems including RIS-RSM, RIS-QAM, and RIS-PSK under equivalent conditions, demonstrating that the proposed systems achieve better error performance. 
\end{abstract}

\begin{IEEEkeywords}
Reconfigurable intelligent surface, spatial modulation, index modulation, Euclidean distance optimized antenna selection, capacity-optimized antenna selection, antenna correlation based antenna selection.
\end{IEEEkeywords}

%
\IEEEpeerreviewmaketitle


\vspace{-0.2cm}
\section{Introduction}\label{sec1}

\IEEEPARstart{T}{he} need for high-performance wireless communication systems is driven by the continuous evolution of user requirements, the massive growth of data traffic, the emergence of new Internet of Things applications, and the anticipation of future network trends. These factors drive the exploration of alternative technologies and the development of radically new communication paradigms to address the wireless communication engineering challenges of the future \cite{6G1, 6G2, 6G3}. These alternative technologies, such as index modulation (IM), new low complexity multiple-input multiple-output (MIMO) schemes, reconfigurable intelligent surface (RIS), terahertz communications, and advanced antenna selection technologies, are being explored to meet the evolving user requirements, new applications, and networking trends anticipated for $2030$ and beyond.

IM techniques have brought a new dimension to wireless communication by offering a new method for information transmission. IM revolutionizes wireless communications by introducing a groundbreaking concept to address the pressing need for exceptional spectral and energy efficiency. IM represents a paradigm shift from traditional digital modulation schemes that rely on manipulating the amplitude/phase/frequency of a sinusoidal carrier signal for transmission, which has dominated the field of communications for the past half-century. By altering the on/off status of transmission entities such as antennas, subcarriers, modulation types, and time slots, IM introduces an entirely new approach to data transmission. Moreover, IM operates in a more energy-efficient manner by selectively deactivating essential system components while still capitalizing on their utility for data transfer. This unique characteristic enables IM to achieve unparalleled energy efficiency while reducing the carbon footprint. Also, the flexible nature of IM empowers wireless communication systems to achieve increased spectral efficiency without adding unnecessary complexity to the hardware \cite{index1,index2,index3,index4, cimsmbm}.

Spatial modulation (SM), one of the most well-known representatives of IM schemes, is an important wireless communication technique that improves the spectral efficiency and error performance of wireless communication systems. SM is a modulation technique that operates by activating only one transmit antenna at any time. The antenna indices serve as an additional source of information, enhancing the spectral efficiency of SM \cite{SM1}. At each transmission time, only one transmit antenna is activated thereby eliminating inter-channel interference (ICI) and the need for antenna synchronization. SM belongs to the single radio frequency (RF) large-scale MIMO wireless systems family, utilizing multiple antennas in a novel way compared to conventional MIMO systems. It maximizes energy efficiency by selectively activating a reduced number of antenna elements, while also achieving high spectral efficiency and reduced complexity. SM conveys additional information bits through the index of the active antenna and achieves a spatial multiplexing gain by exploiting the position of the antenna. Also, it offers benefits such as high energy efficiency, high spectral efficiency, low detection complexity, absence of ICI, no need for inter-antenna synchronization, and compatibility with massive MIMO \cite{SM2, SM3, SM4, SMM}.

The RIS technology, which provides high error performance, stands out among the new wireless communication technologies that have been introduced recently. RISs are artificial surfaces made of electromagnetic material that can manipulate the propagation medium \cite{Salhab2021, RIS01, RIS02}. RIS is a low-cost and low-complexity technology that reduces the energy consumption of wireless networks and improves error performance. RISs do not need an energy source, are not affected by receiver noise, can operate at any operating frequency, are easy to deploy, and do not require converters and power amplifiers \cite{BasarRIS2019, Yuan2021}. RISs are passive devices composed of electromagnetic material that can be easily deployed on various structures. Due to their nearly passive nature and low-cost construction, RISs can be conveniently placed on building facades, indoor walls, aerial platforms, and even incorporated into clothing. Moreover, RISs are environmentally friendly. Unlike conventional relaying systems that require power amplifiers, RISs manipulate incoming signals by controlling the phase shift of each reflecting element. This approach eliminates the need for power amplification and contributes to greater energy efficiency, making RISs an eco-friendly solution for wireless communication \cite{RIS1, RIS2, RIS3, RIS4}. RIS-space shift keying (RIS-SSK) and RIS-SM systems, which are RIS-based IM systems, are presented in \cite{BasarRIS_IM2020}. The proposed RIS-based IM systems provide a high data rate and high error performance. RIS-aided non-orthogonal multiple access (NOMA) network is investigated in \cite{Hou2020}. The results show that the RIS-aided NOMA network outperforms its other orthogonal counterparts. In \cite{RISBURAK}, a novel RIS-aided spatial media-based modulation system, called RIS-SMBM, is proposed as a new approach to enhancing system performance, energy efficiency, and spectral efficiency for Rayleigh fading channels. In the proposed system, SM, MBM, and RIS techniques are combined in a single multi-input single-output (MISO) system.

Antenna selection techniques play a crucial role in improving the performance and efficiency of wireless communication systems. The selection of appropriate antennas significantly improve signal quality, increase capacity, and reduce interference. By choosing the optimal antennas, it becomes possible to mitigate the effects of fading, multipath propagation, and other channel impairments, thereby improving the overall reliability, performance, and robustness \cite{AS1, AS2, CMRC}. Capacity-optimized antenna selection (COAS), antenna correlation antenna selection (ACAS), and Euclidean distance-optimized antenna selection (EDAS) techniques are of paramount importance in improving the performance and capacity of communication systems by optimizing the selection of antennas based on channel characteristics. In the COAS technique, the antenna selection process is carried out passively, where the selected antenna indices set is constituted from the largest channel norms. This approach aims to optimize capacity by selecting antennas that have the highest channel norms, ensuring better signal quality and improved performance in the communication system \cite{fatihcoas, coas2}. The ACAS technique aims to maximize the dissimilarity between selected channels to take advantage of diverse fading conditions. The selection process in the ACAS technique involves evaluating the cosine similarity between each pair of channel vectors and selecting the channels with the highest dissimilarity.  By choosing channels with different fading characteristics, the impact of fading on the transmitted signal can be minimized, leading to improved error performance \cite{burakacas, ACASILK}. The EDAS technique operates passively by selecting antennas based on the maximization of the minimum Euclidean distance (ED) over all possible symbols. This involves determining all possible vector sets and then identifying the specific set of antennas that maximizes the minimum ED. By doing so, the EDAS technique achieves high transmit diversity and improves the overall performance of the system \cite{ERDOGANEDAS, EDAS2}. 

\vspace{-0.2cm}
\subsection{Contributions}

In this paper, new wireless communication systems, COAS-RIS-RSM, ACAS-RIS-RSM, and EDAS-RIS-RSM, are proposed by enhancing the promising RIS-supported RSM scheme with COAS, ACAS, and EDAS antenna selection techniques for 6G and beyond wireless communication networks. The main contributions of the paper can be summarized as follows:
\begin{itemize}

    \item The average bit error rate (ABER) analyses of the proposed AS-RIS-RSM systems are obtained and compared with simulation results for different system parameters. The obtained results show that the analytical and simulation performances are very close to each other.
    \item The capacity analyses of the proposed AS-RIS-RSM systems are derived and compared with RIS-QAM/PSK and RIS-RSM systems. The results obtained show that the proposed systems have higher capacity.
    \item The effect of varying the number of reflecting surface elements and modulation order on the performance of the proposed COAS-RIS-RSM, ACAS-RIS-RSM, and EDAS-RIS-RSM systems is investigated.
    \item In the proposed system, an optimal maximum likelihood (ML) detector and a low-complexity sub-optimal greedy detector (GD) are used to estimate the active receive antenna index and the transmitted symbol.
    \item Computational complexity analyses of proposed COAS-RIS-RSM, ACAS-RIS-RSM, and EDAS-RIS-RSM systems are derived and the results are compared with RIS-PSK, RIS-QAM, and RIS-RSM systems.
    \item The proposed COAS-RIS-RSM, ACAS-RIS-RSM, and EDAS-RIS-RSM systems are compared with RIS-PSK, RIS-QAM, and RIS-RSM systems in terms of error performance over the Rayleigh fading channels and it is shown that the proposed systems provide better results.
  
\end{itemize}

\begin{figure*}[t]
    \centering{\includegraphics[width=0.85 \textwidth]{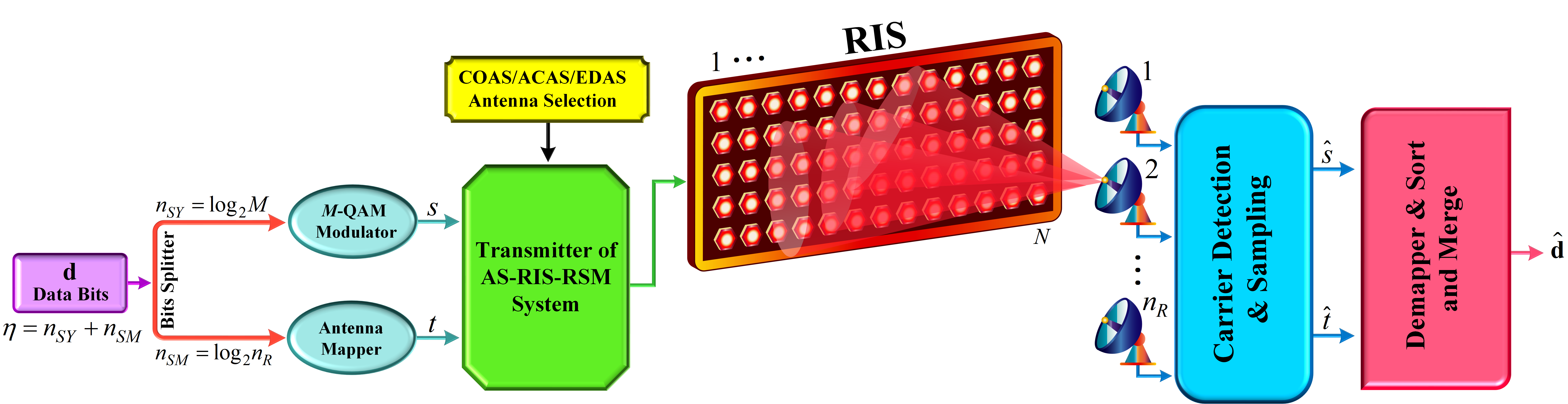}}
    \caption{System model of the proposed AS-RIS-RSM system.}
	\label{SystemModel} 
 \vspace{-0.5cm}
\end{figure*}

\vspace{-0.2cm}
\subsection{Organization and Notations}

The remainder of the article is compartmentalized as follows. In Section II, the AS-RIS-RSM system model and receiver structures are described. Antenna selection techniques for the RIS-RSM scheme are introduced in Section III. The performance analysis of the proposed system is given in Section IV. Section v presents the capacity analyses of the proposed AS-RIS-RSM system. The computational complexity analyses are presented in Section VI. Section VII presents the simulation results containing performance results and discussions of the proposed system and benchmark systems are offered. Finally, conclusions and discussions are offered in Section VIII. 
	
The following notation is used throughout this article. \textit{i}) Bold lower/upper case symbols represent vectors/matrices; $(\cdotp)^T$, $|\cdotp | $, and  $| \cdotp |$ denote transpose, Frobenius norm, and Euclidean norm, respectively, $\Re(.)$ and $\Im(.)$ are the real and imaginary parts of a complex-valued quantity.

\vspace{-0.2cm}
\section{System Model}
Fig. \ref{SystemModel} illustrates the proposed AS-RIS-RSM system model, which operates over the Rayleigh flat fading channel. The proposed system is considered a single-input and multiple-output (SIMO) system. In addition, the proposed system model in Fig. \ref{SystemModel} includes a RIS block containing $N$  reflecting surface elements controlled by a microcontroller, which provides communication between the transmitter and the receiver. These reflecting surface elements efficiently transmit the signal to the selected receiver antenna by rotating the phase of the signal to maximize the signal-to-noise ratio (SNR). Also, since the transmit antenna is deployed very close to the RIS, it is assumed that there is no channel between the transmitter and the RIS \cite{ertoreis}. The proposed system comprises one transmit antenna at the transmitter and $n_R$ receive antennas at the receiver. However, the $n_S$ antennas are selected from the $n_R$ antennas according to COAS, ACAS, and EDAS techniques. When no antenna selection technique is used, all available $n_R$ antennas are used for data transmission and the channel matrix $\textbf{G} \in \mathbb{C}^{N\times n_R}$ is conventionally expressed as follows:
\begin{eqnarray}\label{channel_matrix1}
\textbf{G} &=&\begin{bmatrix}
g_{1,1} & g_{1,2} &  \cdots  & g_{1,n_R}  \\ 
g_{2,1} &g_{2,2} & \cdots  & g_{2,n_R}  \\ 
\vdots & \vdots & \ddots& \vdots\\ 
g_{N,1} & g_{N,2} & \cdots  & g_{N,n_R}
\end{bmatrix}_{N \times n_R}\!\!\!\!\!\!\!\!\!\!\!\!.
\end{eqnarray}
However, according to COAS, ACAS, and EDAS techniques, $n_S$ antennas are selected from $n_R$ receiver antennas and only these $n_S$ selected antennas are used for transmission. Therefore, under the condition $n_S \leq n_R$, only the following channel matrix $\textbf{G}_\mathcal{S} \in \mathbb{C}^{N\times n_S}$ corresponding to the selected $n_S$ antenna is used for data transmission:
\begin{eqnarray}\label{channel_matrix2}
\textbf{G}_\mathcal{S} &=&\begin{bmatrix}
g_{1,1} & g_{1,2} &  \cdots  & g_{1,n_S}  \\ 
g_{2,1} &g_{2,2} & \cdots  & g_{2,n_S}  \\ 
\vdots & \vdots & \ddots& \vdots\\ 
g_{N,1} & g_{N,2} & \cdots  & g_{N,n_S}
\end{bmatrix}_{N \times n_S}\!\!\!\!\!\!\!\!\!\!\!\!.
\end{eqnarray}


In Fig. \ref{SystemModel}, a binary information vector with the size of $1 \times \eta$ denoted as $\textbf{d}$ is to be transmitted using the proposed AS-RIS-RSM system. $\eta=n_{SY}+n_{SM}$ bits are transmitted at a specific time period during the communication process. Furthermore, $\textbf{d}$ is divided into two sub-vectors, one containing $n_{SY}={\log}_2(M)$ bits and the other $n_{SM}={\log}_2(n_{S})$ bits. The bits in the first sub-vector are used to select a symbol from the $M$-QAM symbol space and the bits in the second sub-vector determine the active antenna index.

The received signal transmitted through the reflecting surface elements of the RIS and received by the $\ell^\text{th}$ receive antenna can be expressed as follows:
\begin{eqnarray}\label{receive1}
 y_\ell &=& \sqrt{E_s} \ \Bigg[ \sum^{N}_{r=1}g_{r,\ell} e^{j{\psi_{r}}} \Bigg]\ s_q\ + w_\ell \nonumber  \\
 &=& \ \sqrt{E_s} \ \textbf{g}^{T}_{\ell} \boldsymbol{\Psi} \  s_q + w_\ell,
\end{eqnarray}
where $\textbf{g}_{\ell}\in \mathbb{C}^{N\times 1}$ is the $\ell^{\text{th}}$ column vector of $\textbf{G}_\mathcal{S}$, i.e. $\textbf{G}_\mathcal{S}= [\textbf{g}_{1},\ldots,\textbf{g}_{\ell},\ldots,\textbf{g}_{n_S}]$. The Rayleigh channel coefficient between the $\ell^\text{th}$ receive antenna and the $r^\text{th}$ reflecting surface element is represented as $g_{r,\ell}={\lambda }_{r,\ell} e^{-j{\varphi_{r,\ell}}}$ and $\ell \in \{1, 2, \dots n_S\}$, $r \in \{1,2, \dots N\}$. The $q^{\text{th}}$ complex symbol chosen from the $M$-QAM constellations is expressed as $s_q$.  To maximize the instantaneous SNR at the receiver, the RIS tunes its adjustable phases as $\psi_{r}=\varphi_{r,t}$ according to this selected $t^\text{th}$ antenna. Also, reflection vector $\boldsymbol{\Psi}= \big[ e^{j{{\varphi}_{1,t}}},e^{j{{\varphi}_{2,t}}},\cdots,e^{j{{\varphi}_{N,t}}} \big]^T$. It is evident that the nullification of the channel phase is required to maximize the instantaneous SNR at the receiver. The instantaneous SNR at receiver for the $\ell^\text{th}$ receive antenna is described as follows:
\begin{eqnarray} \label{SNR1} 
\gamma_\ell &=& \frac{\sqrt{E_s}}{N_0} \ \bigg|\sum\limits_{r=1}^{N} \lambda_{r,\ell}  e^{j(\psi_{r}-\varphi_{r,\ell})}\bigg|^2,
\end{eqnarray} 
where to maximize the instantaneous SNR at the selected $t^\text{th}$ receive antenna, $\psi_{r}$ is selected as equal to $\varphi_{r,t}$. As a result, the maximized SNR at the selected $t^\text{th}$ antenna is given as follows:
\begin{eqnarray} \label{SNR2} 
\Gamma_t &=& \frac{\sqrt{E_s}}{N_0} \ \bigg|\sum\limits_{r=1}^{N} \lambda_{r,t} \bigg|^2.
\end{eqnarray}

\subsection{Receiver Structure of the AS-RIS-RSM system}
This section presents the ML and GD used to estimate the symbol and active antenna index in the AS-RIS-RSM receiver. As the system model of the AS-RIS-RSM is shown in Fig. \ref{SystemModel}, the signals from the reflecting surface elements of the RIS are received by the receiving antennas and then the symbol and active antenna index are estimated from the received signals with the help of the carrier detection $\&$ sampling block. This section presents the working principle of the ML and GD of the AS-RIS-RSM system in this carrier detection $\&$ sampling block. Finally, the parameters estimated at the AS-RIS-RSM receiver are used to generate the estimated bit sequence $\hat{\textbf{d}}$, which is estimated using the demapper $\&$ sort and merge block as shown in Fig. \ref{SystemModel}.

\subsubsection{ML Detector for AS-RIS-RSM System}
At the receiver of the AS-RIS-RSM system, the ML detector jointly estimates the active receive antenna index $t$ and the selected symbol index $q$. The ML detector tries all combinations of $t$ and $q$ and assigns the $t$ and $q$ values that give the minimum result as the estimated value. The ML detector of the AS-RIS-RSM system is defined as follows:
\begin{eqnarray}\label{ML}
\Big[\hat{q},\hat{t}\Big] & =& \text{arg}\min_{q,t} \sum^{n_S}_{\ell=1} \Bigg| y_\ell - \Bigg( \sum^{N}_{r=1} g_{r,\ell} e^{j{\varphi_{r,t}}} \Bigg) \ s_q \Bigg|^2  .
\end{eqnarray}
Using the estimated active receive antenna index $\hat{t}$ and the selected symbol index $\hat{q}$, the estimated information bits $\hat{\textbf{d}}$ are obtained with the help of the demapper $\&$ sort and merge block in Fig. \ref{SystemModel}.

\subsubsection{Greedy Detector for AS-RIS-RSM System}
The ML detector appears to be an advantageous estimator with optimal error performance. Although the ML detector has better error performance than other detectors, it has a high computational complexity as it considers all combinations together. Therefore, the GD, which has lower computational complexity than the ML detector, is also included in the receiver structure of the proposed system. The GD does not consider combinations of available symbols and receive antenna indices together as in the ML detector. This results in lower computational complexity. In the GD, the receiving antenna index with the highest instantaneous energy is chosen as the estimated receiving antenna index. Thus, the GD first estimates the active receive antenna index from the received signal as follows:
\begin{eqnarray}\label{greedy1}
\hat{t} & =& \text{arg}\max_{t \in \{1, 2, \dots n_S\}} \Big| y_t \Big|^2  .
\end{eqnarray}
Using the estimated active receive antenna index obtained from ($\ref{greedy1}$), the transmitted symbol index is estimated by the GD as follows:
\begin{eqnarray}\label{greedy2}
\hat{q} & =& \text{arg}\min_{q}  \Bigg| y_{\hat{t}} - \Bigg( \sum^{N}_{r=1} \lambda_{r,\hat{t}} \Bigg) \ s_q \Bigg|^2.
\end{eqnarray}
With the aid of the demapper $\&$ sort and merge block in Fig. \ref{SystemModel}, the estimated information bits $\hat{\textbf{d}}$ are obtained using these estimated active receive antenna index $\hat{t}$ and the selected symbol index $\hat{q}$.

\section{Antenna Selection Techniques for RIS-RSM Scheme}
This section presents the COAS, ACAS, and EDAS techniques used to improve the error performance of the RIS-RSM system.

 \vspace{-0.2cm}

\subsection{COAS-Aided RIS-RSM}

Assume that $n_S$ antennas of the $n_R$ antennas in the receiver side of the proposed AS-RIS-RSM system are selected using the COAS technique. The capacity of the proposed system $\mathcal{T}$ with $n_S$ receive antennas selected from $n_R$ antennas can be bounded as follows:
\begin{eqnarray} \label{bounded} 
\beta \leq \mathcal{T} \leq \beta \ + \ \text{log}_2(n_S),
\end{eqnarray} 
where $\beta= \frac{1}{n_S} \sum_{p=1}^{n_S} \text{log}_2\Big( 1 \ + \ \frac{E_s}{N_0 || \textbf{h}_p ||^2}\Big)$. According to the $\beta$ expression, it is seen that a certain SNR gain is achieved when $n_S$ channel coefficients with the highest norms are selected from $n_R$ receive antennas. Therefore, it is first necessary to calculate the norm of the channel coefficients. The norms of $n_R$ channel coefficients ordered from largest to smallest are expressed as follows \cite{antenilk}:
\begin{eqnarray}\label{coaseq1}
         \big|\big|\mathbf{g}_1\big|\big|^2 \geq \ldots \geq \big|\big|\mathbf{g}_{n_S}\big|\big|^2 \geq \big|\big|\mathbf{g}_{n_{S+1}}\big|\big|^2 \geq \ldots \geq \big|\big|\mathbf{g}_{n_R}\big|\big|^2.
\end{eqnarray}
In (\ref{coaseq1}), the selected channel matrix $\mathbf{G}_{ \mathcal{S}} \in \mathbb{C}^{N\times n_S} $ is given as follows obtained by selecting $n_S$ channel vectors with the highest channel norm among the channel coefficient vectors ordered from smallest to largest:
\begin{eqnarray}\label{coaseq2}
\mathbf{G}_{ \mathcal{S}} = \Big[\textbf{g}_1,\textbf{g}_2, \ldots , \textbf{g}_{n_S}\Big].
\end{eqnarray}

\vspace{-0.2cm}
\subsection{ACAS-Aided RIS-RSM}
Assuming the ACAS approach is employed to select $n_S$ antennas from the available $n_R$ antennas within the receiver side of the proposed AS-RIS-RSM system. The independent positioning of the antennas in the wireless communication system is very important for the performance of the system. Therefore, the ACAS approach is based on selecting a set of antennas that are most distinct (most independent) from each other among the available antennas. ACAS technique uses cosine similarity to select the most distinct sets of antennas. Based on the cosine similarity theory, the similarity between any two channel coefficients $\mathbf{g}_i$ and $\mathbf{g}_j$ is defined as follows:
\begin{eqnarray}\label{cosine_acas}
\text{Similarity}(\mathbf{g}_i, \mathbf{g}_j) &=& \frac{ \big| \mathbf{g}_i^H \mathbf{g}_j \big| } { \big|\big| \mathbf{g}_i  \big|\big| \ \big|\big| \mathbf{g}_j  \big|\big| } .
\end{eqnarray}
where $i,j \in \big\{ 1, 2 \ldots \ n_S \big\}$. With the help of (\ref{cosine_acas}), the selected channel matrix $\mathbf{G}_{ \mathcal{S}} \in \mathbb{C}^{N\times n_S}$ for the AS-RIS-RSM system is expressed as follows \cite{ACASILK}:
\begin{eqnarray}\label{equ_acas}
\mathbf{G}_\mathcal{S} &=& \textrm{arg}\min_{ \textbf{G}_\mathcal{S} \in \mathcal{S}(\textbf{G}_\mathcal{S})}  \Bigg\{\max_{\substack{i,j \\ i \neq j}}  \frac{ \big| \mathbf{g}_i^H \mathbf{g}_j \big| } { \big|\big| \mathbf{g}_i  \big|\big| \ \big|\big| \mathbf{g}_j  \big|\big| }   \Bigg\}.
\end{eqnarray}

\begin{table}[t]
\addtolength{\tabcolsep}{-1pt}
		\caption{Transmission vectors of the AS-RIS-RSM system for $M=4$ and $n_S=4$. (Where the first two bits of the data bits determine the active antenna index and the last two bits determine the active symbol index.)}
\vspace{-1.5 em}	
 \begin{center}
		\label{Tablodata}
		\begin{tabular}{|p{1cm} p{1cm}p{1cm}p{1.45cm}p{1.9cm}|} 
	        \hline
			\hline
		\scriptsize	\textbf{Data Bits}  & \scriptsize \textbf{Active Antenna Index}  & \scriptsize \textbf{Active Symbol Index} & \scriptsize \textbf{Selected $4$-QAM Symbol} & \scriptsize \textbf{Transmission Vector} \\

			\hline
		    $ [0000] $   & $t=1$  & $q = 1$ &  $s_1=-1+j$ & $\textbf{s}=[s_1 \ 0 \ 0\  0]^T $\\  
	        $ [0001] $   & $t=1$  & $q = 2$ &  $s_2=-1-j$  & $\textbf{s}=[s_2 \ 0 \ 0\  0]^T $     \\   
			$ [0010]$    & $t=1$  & $q = 3$ &  $s_3=1+j$  & $\textbf{s}=[s_3 \ 0 \ 0\  0]^T  $  \\
            $ [0011] $   & $t=1$  & $q = 4$ &  $s_4=1-j$  & $\textbf{s}=[s_4 \ 0 \ 0\  0]^T  $  \\

             $ [0100] $  & $t=2$  & $q = 1$ &  $s_1=-1+j$  & $\textbf{s}=[ 0 \  s_1  \ 0\  0]^T $   \\

             $[0101]$  & $t=2$   & $q = 2$ &  $s_2=-1-j$  & $\textbf{s}=[ 0 \  s_2  \ 0\  0]^T $ \\

             $[0110]$  & $t=2$    & $q = 3$ &  $s_3=1+j$  & $\textbf{s}=[ 0 \  s_3  \ 0\  0]^T $   \\

             $[0111]$   & $t=2$    & $q = 4$ &  $s_4=1-j$  & $\textbf{s}=[ 0 \  s_4 \ 0\  0]^T $  \\

             	$ [1000] $  & $t=3$    & $q = 1$ &  $s_1=-1+j$   & $\textbf{s}=[ 0 \ 0 \ s_1 \  0]^T $    \\

	       $ [1001] $ & $t=3$    & $q = 2$ &  $s_2=-1-j$ & $\textbf{s}=[ 0 \ 0 \ s_2 \  0]^T $    \\

			$[1010]$ & $t=3$    & $q = 3$ &  $s_3=1+j$ & $\textbf{s}=[ 0 \ 0 \ s_3 \  0]^T $   \\

              $ [1011] $ & $t=3$       & $q = 4$ &  $s_4=1-j$  & $\textbf{s}=[ 0 \ 0 \ s_4 \  0]^T $  \\

             $ [1100] $  & $t=4$     & $q = 1$ &  $s_1=-1+j$ & $\textbf{s}=[ 0 \ 0 \ 0 \ s_1 ]^T $  \\

             $[1101]$ & $t=4$   & $q = 2$ &  $s_2=-1-j$& $\textbf{s}=[ 0 \ 0 \ 0 \ s_2 ]^T $  \\

             $[1110]$  & $t=4$    & $q = 3$ &  $s_3=1+j$ & $\textbf{s}=[ 0 \ 0 \ 0 \ s_3 ]^T $  \\

             $[1111]$ & $t=4$    & $q = 4$ &  $s_4=1-j$ & $\textbf{s}=[ 0 \ 0 \ 0 \ s_4 ]^T $  
                \\   
			\hline
			\hline
		
		\end{tabular}
	\end{center}
\vspace{-0.3cm}
\end{table}

\subsection{EDAS-Aided RIS-RSM}

Assume that $n_S$ antennas of the $n_R$ antennas in the receiver side of the proposed AS-RIS-RSM system are selected using the EDAS technique. The specific set of receive antennas, which is chosen to maximize the minimum Euclidean distance among all possible transmit vectors within the $N_C=\binom{n_R}{n_S}$ possibilities, is described as follows \cite{antenilk}:
\begin{eqnarray}\label{edas}
\mathcal{S}_{ED} &=& \textrm{arg}\max_{ \mathcal{S} \in \mathcal{C}_i }  \Bigg\{\min_{\substack{\mathbf{s}_1,\mathbf{s}_2\in \mathcal{Q} \\ \mathbf{s}_1 \neq \mathbf{s}_2}}\big|\big|\mathbf{G}_{ \mathcal{S} }\big(\mathbf{s}_{1}-\mathbf{s}_{2}\big)\big|\big|^2\Bigg\},
\end{eqnarray}
where $\textbf{G}_\mathcal{S}$ is the channel matrix containing $n_S$ columns selected by the EDAS technique. $\mathcal{Q}$ represents all possible transmission vectors defined in Table \ref{Tablodata}. Also, $\mathcal{C}_i$ for $i \in \big\{ 1, 2, \ldots N_C \big\}$ is defined as the set encompassing all potential combinations of $N_C$ elements for selecting $n_S$ from $n_R$.

\section{Performance Analysis of the Proposed AS-RIS-RSM System}

In this section, the analysis of the ABER for the RIS-SMBM technique is presented. The data bits transmitted from the transmitter may be inaccurately estimated at the receiver due to disturbances such as reflection, scattering, and diffraction from the Rayleigh fading channel, as well as noise. Therefore, the necessity of error analysis arises. The examination of the pairwise error probability (PEP), which reveals the probability of making incorrect decisions, is significant for error analysis. By utilizing the upper bounding technique, the ABER of the proposed AS-RIS-RSM system can be expressed as follows:
\begin{equation} \label{ABER_eq} 
	\!\! \! \mathcal{P}_{\text{AS-RIS-RSM}} \approx \frac{1}{\eta2^\eta}\sum_{t,\hat{t}}{\sum_{s,\hat{s}}{\mathcal{P}\big( t,s \rightarrow \hat{t},\hat{s} \big)\mspace{2mu} \mathcal{E} \big( {t,s \rightarrow \hat{t},\hat{s}} \big) }}, 
\end{equation} 
where $\mathcal{P}\big( t,s \rightarrow \hat{t},\hat{s} \big)$ is represented as the average PEP and $\mathcal{E} \big( {t,s \rightarrow \hat{t},\hat{s}} \big)$ is the number of bit errors associated with the pairwise error event. As seen in (\ref{ABER_eq}), in order to calculate the ABER of the proposed AS-RIS-RSM, the PEP expression symbolized by $\mathcal{P}\big( t,s \rightarrow \hat{t},\hat{s} \big)$ is needed. The conditional PEP (CPEP) expression $\mathcal{P}\big( t,s \rightarrow \hat{t},\hat{s}\big|  \mathcal{H} \big) $ associated with the active $t^{\text{th}}$ receiver antenna index and the selected symbol determined according to the data bits can be defined as: 
\begin{eqnarray} \label{PEP_eq1}
  \mathcal{P}\big( t,s \rightarrow \hat{t},\hat{s}\big|  \mathcal{H} \big)   &\!\!\!\!\!=&\! \!\!\!\!\!\mathcal{P} \Bigg( \sum^{n_S}_{\ell=1}  \big| y_\ell -\mathcal{H}_\ell s \big|^2 \!> \!\sum^{n_S}_{\ell=1} \big| y_\ell - \hat{\mathcal{H}}_\ell \hat{s} \big|^2 \Bigg)  \nonumber\\
     &\!\!\!\!\!\!\!\!\!\!\!\!\!\!\!\!\!\!\!\!\!\!\!\!\!\!\!\!\!\!\!\!\!\!\!\!\!\!\!\!\!\!\!\!\!\!\!\!\!\!\!\!\!\!=& \!\!\!\!\!\!\!\!\!\!\!\!\!\!\!\!\!\!\!\!\!\!\!\!\!\!\!\!\! \small  \mathcal{P} \! \Bigg(  \sum^{n_S}_{\ell=1} \!\! -  \big| \mathcal{H}_\ell s - \hat{\mathcal{H}}_\ell \hat{s} \big|^2 \!\! - \!2\Re \big\{ w^*\big( \mathcal{H}_\ell s - \hat{\mathcal{H}}_\ell \hat{s}  \big) \big\} \!\!> \!0 \Bigg) \nonumber  \\ 
   &\!\!\!\!\!\!\!\!\!\!\!\!\!\!\!\!\!\!\!\!\!\!\!\!\!\!\!\!\!\!\!\!\!\!\!\!\!\!\!\!\!\!\!\!\!\!\!\!\!\!\!\!\!\!=&\!\!\!\!\!\!\!\! \!\!\!\!\!\!\!\!\!\!\!\!\!\!\!\!\!\!\!\!\! \mathcal{P} \big( \mathcal{K}  > 0 \big),
\end{eqnarray}
where $\mathcal{H}_\ell$ and $\hat{\mathcal{H}}_\ell$ expressions are defined as $\mathcal{H}_\ell=\sum^{N}_{r=1} g_{r,\ell} e^{j{\varphi_{r,t}}}$ and $\hat{\mathcal{H}}_\ell=\sum^{N}_{r=1} g_{r,\ell} e^{j{\varphi_{r,\hat{t}}}}$.
Also, $\mathcal{K} \sim \mathcal{N}(\mu_\mathcal{D},\sigma^2_\mathcal{D})$ is Gaussian random variable and $\mu_\mathcal{K}$ and $\sigma^2_\mathcal{K}$ are expressed as follows, respectively:
\begin{eqnarray}\label{ortalama}
  \!\!\mu_\mathcal{K} \!\! \!\!&=& \!\!\!\!E\Bigg\{ \sum^{n_S}_{\ell=1} - \big| \mathcal{H}_\ell s - \hat{\mathcal{H}}_\ell \hat{s} \big|^2  - 2\Re \big\{ w^*\big( \mathcal{H}_\ell s - \hat{\mathcal{H}}_\ell \hat{s} \big\} \Bigg\} \nonumber\\ 
  &\!\!\!\!\!\!=&\!\!\!\!\!\! - E \Bigg\{ \sum^{n_S}_{\ell=1} \big| \mathcal{H}_\ell s - \hat{\mathcal{H}}_\ell \hat{s} \big|^2 \Bigg\} = - \sum^{n_S}_{\ell=1} \big| \mathcal{H}_\ell s - \hat{\mathcal{H}}_\ell \hat{s} \big|^2\!\!,
\end{eqnarray}
\begin{eqnarray}\label{varyans}
  \sigma^2_{\mathcal{K}} \!\!\!\! &=& \!\!\!\! \text{Var} \Bigg\{ \sum^{n_S}_{\ell=1} - \big| \mathcal{H}_\ell s - \hat{\mathcal{H}}_\ell \hat{s} \big|^2  - 2\Re \big\{ w^*\big( \mathcal{H}_\ell s - \hat{\mathcal{H}}_\ell \hat{s} \big) \big\}  \Bigg\} \nonumber\\ 
  &\!\!\!\!\!\!\!\!\!\!\!\!=&\!\!\!\! \!\!\!\!\!\! 4 \!\sum^{n_S}_{\ell=1} \big| \mathcal{H}_\ell s - \hat{\mathcal{H}}_\ell \hat{s} \big|^2 \text{Var}\big( w^* \big)=2N_0 \!\! \sum^{n_S}_{\ell=1} \!\big| \mathcal{H}_\ell s - \hat{\mathcal{H}}_\ell \hat{s} \big|^2\!\!. 
\end{eqnarray} 
As a result, using the $Q$ function defined as $\mathcal{P}\left(Z>z\right) = Q\left(z\right) = \frac{1}{2\pi}\int_{z}^{\infty}\exp\left({\frac{-x^2}{2}}\right)dx$ \cite{simon2005digital}, the PEP expression is obtained again as follows:
\begin{eqnarray}\label{eq_cpep}
\mathcal{P}\big( t,s \rightarrow \hat{t},\hat{s}\big|  \mathcal{H} \big)   &=&   \mathcal{P} \big( \mathcal{K}  > 0 \big)  =  Q\left(\frac{-\mu_\mathcal{K}}{\sigma_\mathcal{K}}\right) \nonumber \\ 
&=&  Q  \left(  \sqrt{\frac{\sum^{n_S}_{\ell=1} \big| \mathcal{H}_\ell s - \hat{\mathcal{H}}_\ell \hat{s} \big|^2}{2N_0}}  \right) \nonumber \\    &=&  Q  \left(  \sqrt{\frac{\Upsilon}{2N_0}}  \right), 
\end{eqnarray} 
where, $\Upsilon$ is defined as $\Upsilon \triangleq \sum^{n_S}_{\ell=1} \big| \mathcal{H}_\ell s - \hat{\mathcal{H}}_\ell \hat{s} \big|^2$. Depending on whether the receive antenna index $t$ is detected correctly or incorrectly, two cases occur. The average PEP expression is obtained using the following two cases.

\emph{Case 1} $\left( t\neq\hat{t} \right)$: With the assumption that $\Upsilon \triangleq \Upsilon_{1}+\Upsilon_{2}+\Upsilon_{3}$, the $\Upsilon_{1}$, $\Upsilon_{2}$, and $\Upsilon_{3}$ is given as follows:
\begin{eqnarray}\label{esitdglups}
    \Upsilon_{1} &=& \big| \mathcal{H}_t s - \hat{\mathcal{H}}_t \hat{s} \big|^2 =  \Bigg|  s \sum^{N}_{r=1} {\lambda }_{r,t} - \hat{s} \sum^{N}_{r=1} g_{r,t} \ e^{j{\varphi_{r,\hat{t}}}} \Bigg|^2, \nonumber \\
      \Upsilon_{2} &=& \big| \mathcal{H}_{\hat{t}} s - \hat{\mathcal{H}}_{\hat{t}} \hat{s} \big|^2 =  \Bigg|  s \sum^{N}_{r=1} g_{r,\hat{t}} \ e^{j{\varphi_{r,t}}} - \hat{s} \sum^{N}_{r=1} {\lambda }_{r,\hat{t}}  \Bigg|^2, \nonumber \\
      \Upsilon_{3} &=& \sum^{n_S}_{\ell=1 \left( \ell \neq t, \ell \neq \hat{t} \right) }  \big| \mathcal{H}_\ell s - \hat{\mathcal{H}}_\ell \hat{s} \big|^2,
\end{eqnarray}
where $\Upsilon_{1}$, $\Upsilon_{2}$, and $\Upsilon_{3}$ are obtained for cases $\left(\ell=t\right)$, $\left(\ell = \hat{t}\right)$, and $\left(\ell \neq t, \ell \neq \hat{t}\right)$, respectively.

\emph{Case 2} $\left( t = \hat{t} \right)$: This results in $\mathcal{H}_\ell=\hat{\mathcal{H}}_{\ell}$. The $\Upsilon$ expression for this case can be defined as follows:
\begin{eqnarray}\label{esitups}
    \Upsilon  &=&  \sum^{n_S}_{\ell=1} \big| \mathcal{H}_\ell (s - \hat{s}) \big|^2.
\end{eqnarray}
Consequently, using the $\Upsilon$ expressions defined in (\ref{esitdglups}) and (\ref{esitups}), the average PEP expression is calculated as follows:
\begin{eqnarray}\label{eq_apep}
\mathcal{P}\big( t,s \rightarrow \hat{t},\hat{s} \big)  &=& E_{\mathcal{H}} \bigg\{ \mathcal{P}\big( t,s \rightarrow \hat{t},\hat{s}\big|  \mathcal{H} \big)  \bigg\}, \nonumber \\ 
&=& E_{\mathcal{H}} \left\{ Q  \left(  \sqrt{\frac{\sum^{n_S}_{\ell=1} \big| \mathcal{H}_\ell s - \hat{\mathcal{H}}_\ell \hat{s} \big|^2}{2N_0}} \right) \!\! \right\}, \nonumber \\ 
&=& E_{\mathcal{H}} \Bigg\{Q\Bigg(  \sqrt{\frac{\Upsilon}{2N_0}} \Bigg)\Bigg\}.  
\end{eqnarray}

Finally, by substituting (\ref{eq_apep}) into the (\ref{ABER_eq}), we obtain the ABER of the proposed AS-RIS-RSM system.

\section{Capacity Analyses}

In this section, the capacity analysis of the proposed AS-RIS-RSM system is derived. The ergodic capacity of a traditional MIMO system is expressed as follows \cite{capacity}:
\begin{eqnarray} 
\label{capa1}
\mathcal{C} = E_{\mathbf{G}} \Big\{ \underset{P_{s_q}}{\text{max}}  \ I\big(s_q;\textbf{y} \: | \: \mathbf{G}\big) \Big\},
\end{eqnarray}
$I\big(s_q;\textbf{y} \: | \: \mathbf{G}\big)$ denotes the mutual information between the transmitted vector $s_q$ and the received vector $\textbf{y} = \big[y_1,y_2, \ldots , y_{n_S}\big]^T\in \mathbb{C}^{n_S \times 1}$ under the condition of a known $\mathbf{G}$. In the proposed AS-RIS-RSM system, unlike the traditional MIMO system, the receiving antenna indices are also used for the transmission of data bits. Therefore, the capacity of the proposed AS-RIS-RSM system is obtained by maximizing the mutual information between the $M$-QAM symbol and the active receive antenna indices with $\textbf{y}$. As a result, inspired by \cite{anacapa}, the capacity of the proposed AS-RIS-RSM system is expressed as follows:
\begin{eqnarray} 
\label{capa2}
\mathcal{C} =  E_{\mathbf{G}_\mathcal{S}} \Big\{\underset{\mathcal{P}_{\textbf{v}_\ell}}{\text{max}}  \ I\big(\textbf{v}_\ell, s_q;\textbf{y} \big) \Big\},
\end{eqnarray}
where $\textbf{v}_\ell \in \mathbb{C}^{n_S \times 1 } \triangleq {\big[ 0\, \,0\cdots 0, \,\underset{\stackrel{\uparrow}{ \ell^{\text{th}} \  \text{index} }} 1, 0\, \,\cdots 0\, \,0\big]^T}$ is a vector with $\ell^{\text{th}}$ element $1$ and the other elements are $0$. The received signal in vector-matrix form can be redefined as:
\begin{figure}[t]
\centering{\includegraphics[width=0.48\textwidth]{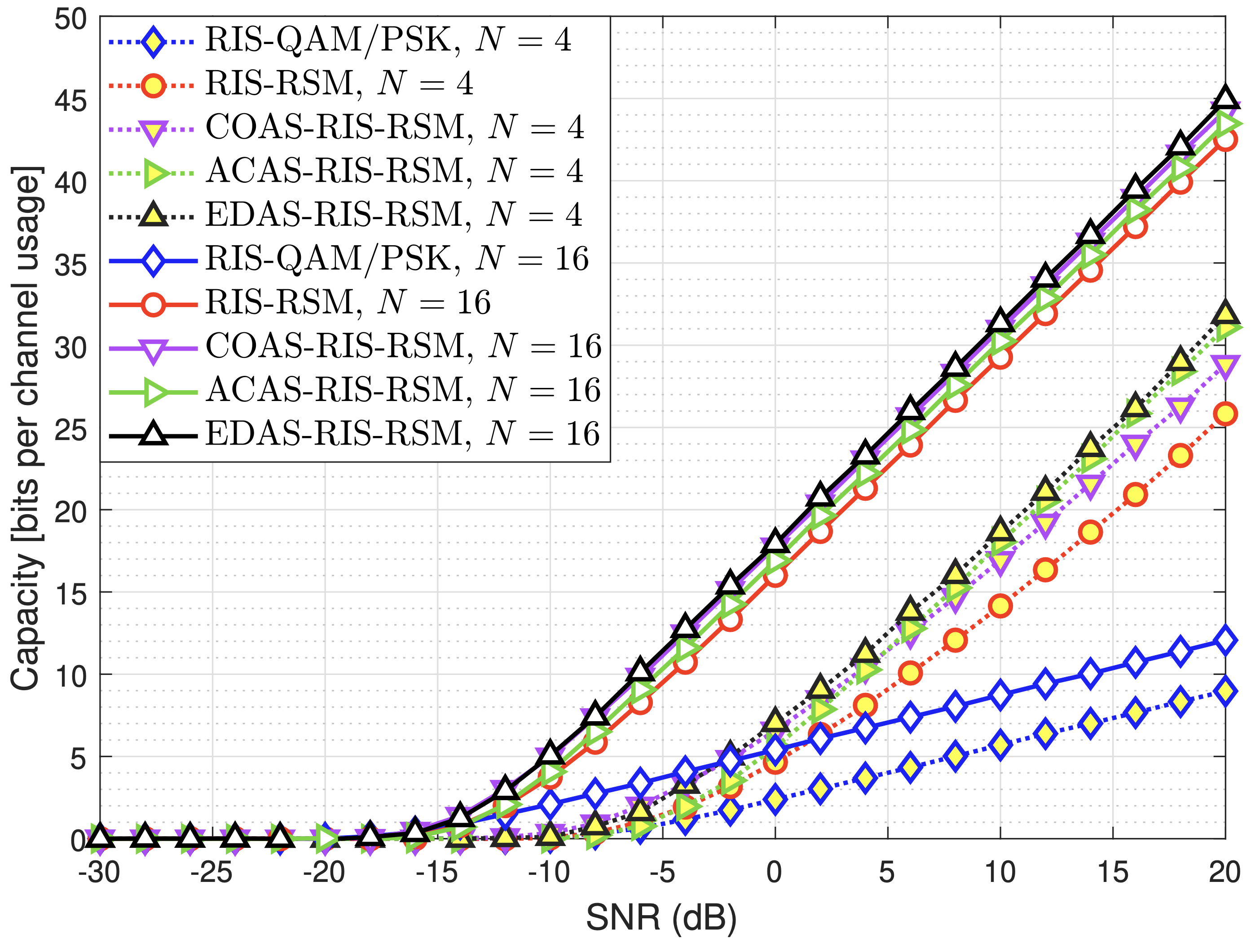}}
	\caption{Capacity comparisons of the RIS, RIS-RSM, COAS-RIS-RSM, ACAS-RIS-RSM and EDAS-RIS-RSM systems for $n_R=16$, $n_S=4$, $M=4$.}
	\label{AS-RIS-RSM-capacity} 
\end{figure}
\begin{eqnarray} \label{received_signal2} 
\textbf{y} &=& \sqrt{E_s} \: \textbf{I}_{n_S}\: \textbf{G}_\mathcal{S}^T  \:\boldsymbol{\Phi}\:\textbf{v}_\ell  \:  s_q\ + \textbf{w}
\end{eqnarray} 
where $\textbf{I}_{n_S}$ is the identity matrix with $n_S$ dimensions.  $\boldsymbol{\Phi}$ refers to the entire phase matrix containing phase information between receive antennas and RIS elements. The mutual information $I\big(\textbf{v}_\ell, s_q;\textbf{y} \big)$ corresponding to the accurately estimated number of bits at the receiver in (\ref{capa2}) is expressed as follows:
\begin{eqnarray} 
\label{capa3}
I\big(\textbf{v}_\ell, s_q;\textbf{y} \big) = \textsc{H}(\textbf{y}) - \textsc{H}\Big(\textbf{y} \ \big| \ \textbf{v}_\ell, s_q \Big),
\end{eqnarray}
where $\textsc{H}(.)$ is defined as the entropy function. The entropy of $\textbf{y}$ when $\textbf{v}_\ell$ and $s_q$ are known is defined as follows:
\begin{eqnarray} 
\label{capa4}
\textsc{H}\Big(\textbf{y} \ \big| \ \textbf{v}_\ell, s_q \Big) = \textsc{H}(\textbf{w}) = \text{log}_2 \big| \pi e N_0 \textbf{I}_{n_S} \big|,
\end{eqnarray} 
where $e\sim\mathcal{CN}(0, N_0\textbf{I}_{n_S})$ and $\textbf{y}\sim\mathcal{CN}(0, \sigma^2_\textbf{y}\textbf{I}_{n_S})$ have complex Gaussian distribution. Also, in (\ref{capa4}), it is seen that in order to maximize the mutual information $I\big(\textbf{v}_\ell, s_q;\textbf{y} \big)$ of the proposed AS-RIS-RSM system, it is necessary to maximize the $\textsc{H}(\textbf{y})$. As a result, $\textsc{H}(\textbf{y})$ is expressed as follows:
\begin{eqnarray} 
\label{capa6}
\textsc{H}(\textbf{y}) = -E_\textbf{y} \big\{ \text{log}_2 \mathcal{P}(\textbf{y})\big\} = \text{log}_2 \big| \pi e \textbf{R}_{\textbf{y}\textbf{y}} \big|,
\end{eqnarray} 
where $\textbf{R}_{\textbf{y}\textbf{y}}$ stand for covariance matrix of 
the $\textbf{I}_{n_S}\textbf{G}_\mathcal{S}^T\boldsymbol{\Phi}\textbf{v}_\ell   s_q$ and is defined as follows:
\begin{eqnarray} 
\label{capa7}
\textbf{R}_{\textbf{y}\textbf{y}} &=& E(\textbf{y}\textbf{y}^H \big| \textbf{v}_\ell ) \nonumber \\ 
&=&\frac{1}{n_S} \sum^{n_S}_{\ell=1}   \textbf{G}_\mathcal{S}^T\boldsymbol{\Phi}\textbf{v}_\ell\textbf{v}_\ell^H\boldsymbol{\Phi}^H\textbf{G}_\mathcal{S}^*+ N_0 \textbf{I}_{n_S}.
\end{eqnarray}
Eventually, the capacity of the proposed AS-RIS-RSM system is obtained as follows:
\begin{eqnarray} 
\label{capa_son}
\mathcal{C} &\!\!\!\!=&\!\!\! \! E_{\textbf{G}_\mathcal{S}} \Bigg\{\! \text{log}_2 \text{det}  \Bigg( \! \! \textbf{I}_{n_S} \!+ \!\frac{\sum^{n_S}_{\ell=1} \textbf{G}_\mathcal{S}^T\boldsymbol{\Phi}\textbf{v}_\ell\textbf{v}_\ell^H\boldsymbol{\Phi}^H\textbf{G}_\mathcal{S}^*}{ n_S N_0}  \Bigg) \!\! \Bigg\}\!.
\end{eqnarray} 
Fig. \ref{AS-RIS-RSM-capacity} presents capacity comparisons of RIS-QAM/PSK, RIS-RSM, COAS-RIS-RSM, ACAS-RIS-RSM, and EDAS-RIS-RSM systems for various system parameters. Capacity comparisons in Fig. \ref{AS-RIS-RSM-capacity} are performed for system parameters $n_R=16$, $n_S=4$, $M=4$, and $N=4, 16$. Firstly, it is clearly seen that the capacities of all systems increase as the number of reflecting surface elements $N$ increases. It is also observed that the proposed COAS-RIS-RSM, ACAS-RIS-RSM, and EDAS-RIS-RSM systems have higher capacity than the compared RIS-QAM/PSK and RSM systems for all cases.

\begin{table}[t]

		\caption{Computational Complexities of RIS-QAM, RIS-PSK, RIS-RSM, COAS-RIS-RSM, ACAS-RIS-RSM, and EDAS-RIS-RSM Systems for $\eta$ bits.}
	
	\begin{center}
		\label{Tablocomplexity}
		\begin{tabular}{|c|c|} 
		 
	        \hline
                \hline
			  RIS-aided Systems & Real Multiplications (RMs) \\ 	
		
			\hline
			
			\!\!$\mathcal{O}_{\text{ COAS-RIS-RSM-ML }}$\!\!  &  $4Nn_R + (N+M)n^2_S$  \\   
				
				\!\!$\mathcal{O}_{\text{ ACAS-RIS-RSM-ML }}$\!\!  &  $(12N+1)\binom{n_R}{n_S} \binom{n_S}{2} + (N+M)n^2_S$  \\   
		
			$\mathcal{O}_{\text{EDAS-RIS-RSM-ML}}$  &               $4N(M-1)\binom{n_R}{n_S} \binom{n_S}{2} + (N+M)n^2_S$ \\  
			
			$\mathcal{O}_{\text{ RIS-RSM-ML}}$   &  $(N+M)n^2_S$  \\ 
		
            $\mathcal{O}_{\text{RIS-QAM/PSK-ML}}$ &  $\Big[ N+M \Big] \bigg(1+\dfrac{{\log}_2(n_{S})}{{\log}_2(M)}\bigg)$ \\ 

            $\mathcal{O}_{\text{ RIS-RSM-Greedy}}$   &  $M+n_S$  \\ 
            
            \!\!$\mathcal{O}_{\text{ COAS-RIS-RSM-Greedy }}$\!\!  &  $4Nn_R + (M+n_S)$  \\ 

            \!\!$\mathcal{O}_{\text{ ACAS-RIS-RSM-Greedy }}$\!\!  &  $(12N+1)\binom{n_R}{n_S} \binom{n_S}{2} + (M+n_S)$  \\ 

            $\mathcal{O}_{\text{EDAS-RIS-RSM-Greedy}}$  &               $4N(M-1)\binom{n_R}{n_S} \binom{n_S}{2} + (M+n_S)$ \\  
            
			\hline
			\hline

		\end{tabular}
	\end{center}

\end{table}

\section{Computational Complexity Analyses}

This section presents computational complexity analyses of the proposed system with COAS, ACAS, and EDAS techniques. Computational complexity analyses in this section are performed in terms of real multiplications (RMs). First, some information for the complexity analysis is considered. The Frobenius norm operation $|| a+jb ||^2$ involves in $4$ RMs. Where $a$ and $b$ are any two real numbers. Also, the multiplication of two complex numbers requires $4$ RMs. 

\begin{figure}[t]
\centering
\includegraphics[scale=0.19]{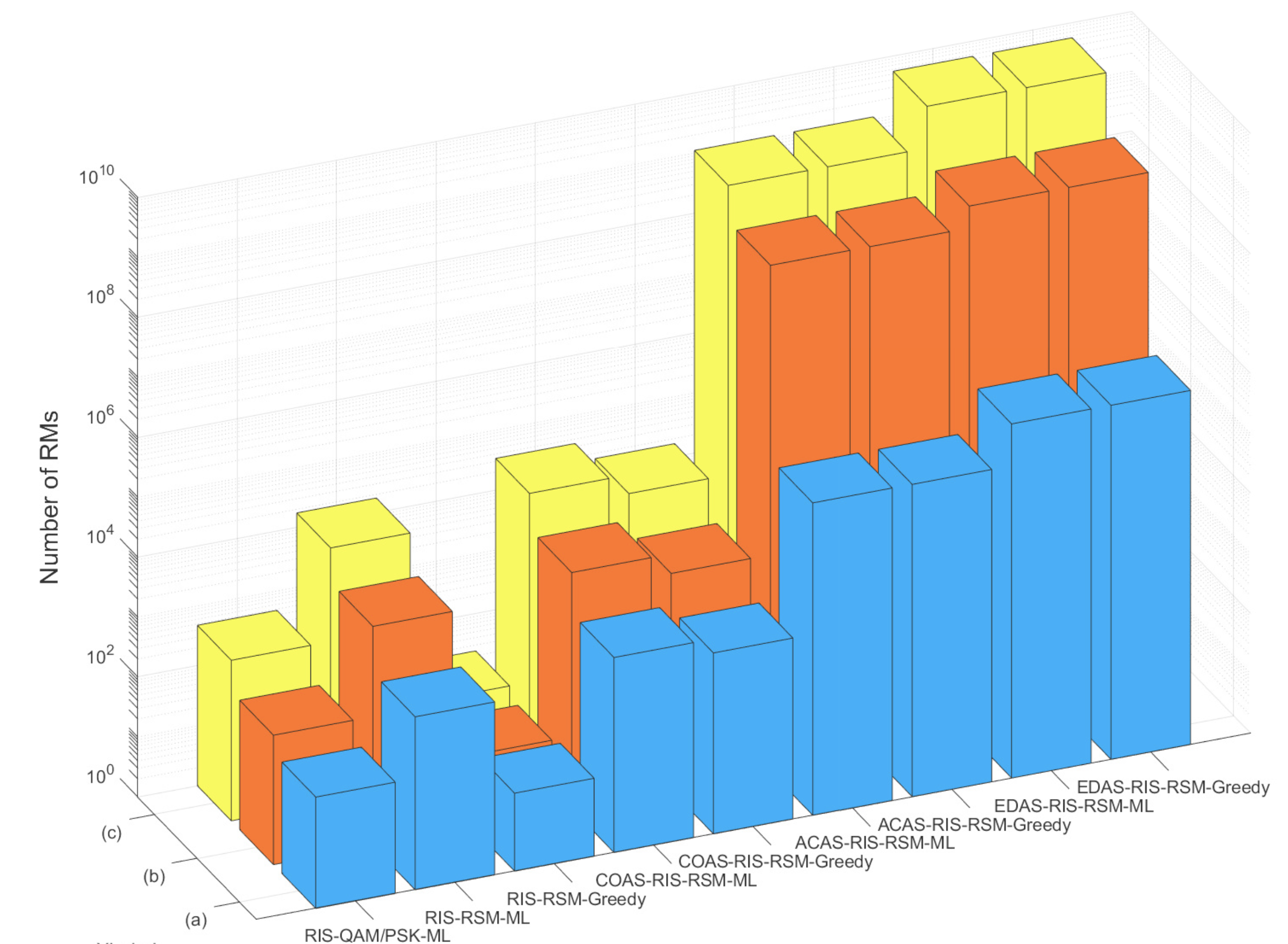}
 \caption{Numerical Examples of The Computational Complexity Comparisons of RIS-QAM, RIS-PSK, RIS-RSM, COAS-RIS-RSM, ACAS-RIS-RSM, and EDAS-RIS-RSM systems for $\eta$ bits while $M=16$, $n_R=8$, $n_S=4$, $N=32$  for (a), $M=8$, $n_R=16$, $n_S=8$, $N=64$  for (b), and $M=16$, $n_S=8$, $n_R=16$, $N=256$ for (c).}
\label{complx_se}
\end{figure}

\begin{table}[t]

		\caption{Numerical Examples of The Computational Complexities in RMs of RIS-QAM, RIS-PSK, RIS-RSM, COAS-RIS-RSM, ACAS-RIS-RSM, and EDAS-RIS-RSM Systems for $\eta$ bits while $M=16$, $n_R=8$, $n_S=4$, $N=32$  for (a), $M=8$, $n_R=16$, $n_S=8$, $N=64$  for (b), and $M=16$, $n_S=8$, $n_R=16$, $N=256$ for (c).}
	
	\begin{center}
		\label{NumTablocomplexity}
		\begin{tabular}{|c|c|c|c|} 
		 
	        \hline
                \hline
			  RIS-aided Systems & (a) & (b) & (c) \\
		
			\hline
			
			\!\!$\mathcal{O}_{\text{RIS-QAM/PSK-ML}}$\!\!  &  $72$ & $144$  & $476$ \\   
				
				\!\!$\mathcal{O}_{\text{ RIS-RSM-ML}}$   & $768$ & $4608$ & $17408$ \\   
		
			$\mathcal{O}_{\text{ RIS-RSM-Greedy}}$   &  $20$  & $16$ & $24$ \\  
			
			$\mathcal{O}_{\text{ COAS-RIS-RSM-ML}}$   &  $1792$ & $8704$ & $33792$  \\ 
		
            $\mathcal{O}_{\text{ COAS-RIS-RSM-Greedy}}$  &  $1044$ & $4112$ & $16408$ \\ 

            $\mathcal{O}_{\text{ACAS-RIS-RSM-ML}}$   &  $162468$ & $277121448$ & $1107403688$ \\ 
            
            \!\!$\mathcal{O}_{\text{ ACAS-RIS-RSM-Greedy }}$\!\!  &  $161720$  & $277116856$ & $1107386304$ \\ 

            \!\!$\mathcal{O}_{\text{ EDAS-RIS-RSM-ML }}$\!\!  &  $807168$  & $645769728$ & $5535147008$ \\ 

            $\mathcal{O}_{\text{EDAS-RIS-RSM-Greedy}}$  &  $806420$ & $645765136$ &  $5535129624$ \\  
            
			\hline
			\hline

		\end{tabular}
	\end{center}

\end{table}

\subsection{Computational Complexity Analysis for COAS-RIS-RSM System}

The Frobenius norm operator in (\ref{coaseq1}) creates $4N$ RMs and this operator is repeated $n_R$ times. As a result, the computational complexity for the COAS technique is obtained as $\mathcal{O}_{\text{COAS}}=4Nn_R$ RMs. Finally, the computational complexity of the COAS-RIS-RSM system can be given as $\mathcal{O}_{\text{COAS-RIS-RSM-ML}}=\mathcal{O}_{\text{COAS}}+\mathcal{O}_{\text{RIS-RSM-ML}}=4Nn_R + (N+M)n^2_S$ RMs and $\mathcal{O}_{\text{COAS-RIS-RSM-Greedy}}=\mathcal{O}_{\text{COAS}}+\mathcal{O}_{\text{RIS-RSM-Greedy}}=4Nn_R + (M+n_S)$ RMs.

\subsection{Computational Complexity Analysis for ACAS-RIS-RSM System}

Each $n_S$ antenna is selected from the available $n_R$ antennas and two antenna sets are selected from the $n_S$ antennas and compared. Therefore, the operation $\frac{ \big| \mathbf{g}_i^H \mathbf{g}_j \big| } { \big|\big| \mathbf{g}_i  \big|\big| \ \big|\big| \mathbf{g}_j  \big|\big| }$ in (\ref{equ_acas}) is repeated $\binom{n_R}{n_S} \binom{n_S}{2}$ times. $\big| \mathbf{g}_i^H \mathbf{g}_j \big|$, $\mathbf{g}_i$, and $\mathbf{g}_j$ each require $4N$ RMs, separately. Also, the division operation also requires $1$ RMs. Consequently, the computational complexity of the ACAS technique is $\mathcal{O}_{\text{ACAS}}=(12N+1)\binom{n_R}{n_S} \binom{n_S}{2}$ RMs. Finally, the computational complexity of the COAS-RIS-RSM system can be given as $\mathcal{O}_{\text{ACAS-RIS-RSM-ML}}=\mathcal{O}_{\text{ACAS}}+\mathcal{O}_{\text{RIS-RSM-ML}}=(12N+1)\binom{n_R}{n_S} \binom{n_S}{2} + (N+M)n^2_S$ RMs and $\mathcal{O}_{\text{ACAS-RIS-RSM-Greedy}}=\mathcal{O}_{\text{ACAS}}+\mathcal{O}_{\text{RIS-RSM-Greedy}}=(12N+1)\binom{n_R}{n_S} \binom{n_S}{2} + (M+n_S)$ RMs.

\subsection{Computational Complexity Analysis for EDAS-RIS-RSM System}

Similar to ACAS, the EDAS technique first selects $n_S$ antenna clusters from $n_R$ available antenna clusters. Then, two symbols are selected from the symbol space, and two antennas are selected from $n_S$ antennas. Therefore, the computational complexity of the EDAS technique is expressed as $\mathcal{O}_{\text{EDAS}}=4N(M-1)\binom{n_R}{n_S} \binom{n_S}{2}$. Finally, the computational complexity of the EDAS-RIS-RSM system can be given as $\mathcal{O}_{\text{EDAS-RIS-RSM-ML}}=\mathcal{O}_{\text{EDAS}}+\mathcal{O}_{\text{RIS-RSM-ML}}= 4N(M-1)\binom{n_R}{n_S} \binom{n_S}{2} + (N+M)n^2_S$ RMs and $\mathcal{O}_{\text{EDAS-RIS-RSM-Greedy}}=\mathcal{O}_{\text{EDAS}}+\mathcal{O}_{\text{RIS-RSM-Greedy}}= 4N(M-1)\binom{n_R}{n_S} \binom{n_S}{2} + (M+n_S)$ RMs.

Furthermore, the computational complexity of the RIS-RSM system is given as $\mathcal{O}_{\text{RIS-RSM-ML}}=(N+M)n^2_S$ for ML detector and $\mathcal{O}_{\text{RIS-RSM-Greedy}}=(M+n_S)$ for GD in \cite{BasarRIS_IM2020}. Similarly, the computational complexity of RIS-QAM and RIS-PSK systems is $\mathcal{O}_{\text{RIS-QAM-ML}}=\mathcal{O}_{\text{RIS-PSK-ML}}=N+M$ for ML detector. The computational complexities of RIS-QAM, RIS-PSK, RIS-RSM, COAS-RIS-RSM, ACAS-RIS-RSM, and EDAS-RIS-RSM systems are given depending on the system parameters in Table \ref{Tablocomplexity}. Also, Fig. \ref{complx_se} presents computational complexity comparisons of RIS-QAM, RIS-PSK, RIS-RSM, COAS-RIS-RSM, ACAS-RIS-RSM, and EDAS-RIS-RSM systems. The system parameters $M=16$, $n_R=8$, $n_S=4$, $N=32$ are used for Fig. \ref{complx_se} (a), $M=8$, $n_R=16$, $n_S=8$, $N=64$ for Fig. \ref{complx_se} (b), and $M=16$, $n_S=8$, $n_R=16$, $N=256$ for Fig. \ref{complx_se} (c), respectively. For RIS-RSM, COAS-RIS-RSM, ACAS-RIS-RSM, and EDAS-RIS-RSM systems, computational complexity results are presented separately for the cases where ML and GD are used. It can be seen that using the GD results in lower computational complexity than the ML detector. However, since the antenna selection complexity of ACAS and EDAS techniques is very high, the computational complexity is not sufficiently reduced when using the GD for the ACAS-RIS-RSM and EDAS-RIS-RSM systems. Fig. \ref{complx_se} shows that the EDAS-RIS-RSM system has the highest computational complexity and the RIS-QAM and RIS-PSK systems have the lowest computational complexity. However, it is emphasized that the EDAS-RIS-RSM system has the best BER performance and the RIS-QAM and RIS-PSK systems have the worst BER performance. Therefore, it is comprehended that as the error performance improves, computational complexity increases. Also, the computational complexity values of RIS-QAM, RIS-PSK, RIS-RSM, COAS-RIS-RSM, ACAS-RIS-RSM, and EDAS-RIS-RSM systems are presented in RMs in Table \ref{NumTablocomplexity} for clearer seeing of the computational complexity values presented in Fig. \ref{complx_se}.

\section{Simulation Results}
In this section, simulation results of the proposed RSM-RIS-AS system are presented and the results are discussed in comparison with its counterpart wireless communication systems. The impact of various system parameters on the error performance of the proposed system, which includes RSM-RIS-COAS, RSM-RIS-ACAS, and RSM-RIS-EDAS systems, and are called RSM-RIS-AS for short, is investigated. The proposed RSM-RIS-COAS, RSM-RIS-ACAS, and RSM-RIS-EDAS systems have been compared with RIS-PSK, RIS-QAM, and RIS-RSM systems in terms of error performance. The simulation results are obtained for the $M$-QAM technique in Rayleigh fading channels. In all simulations, SNR value is expressed as $\mathrm{SNR (dB)} = 10\log_{10}(E_s/N_0)$, where $E_s$ represents the transmitted signal energy.

\begin{figure}[t]
\centering{\includegraphics[width=0.48\textwidth]{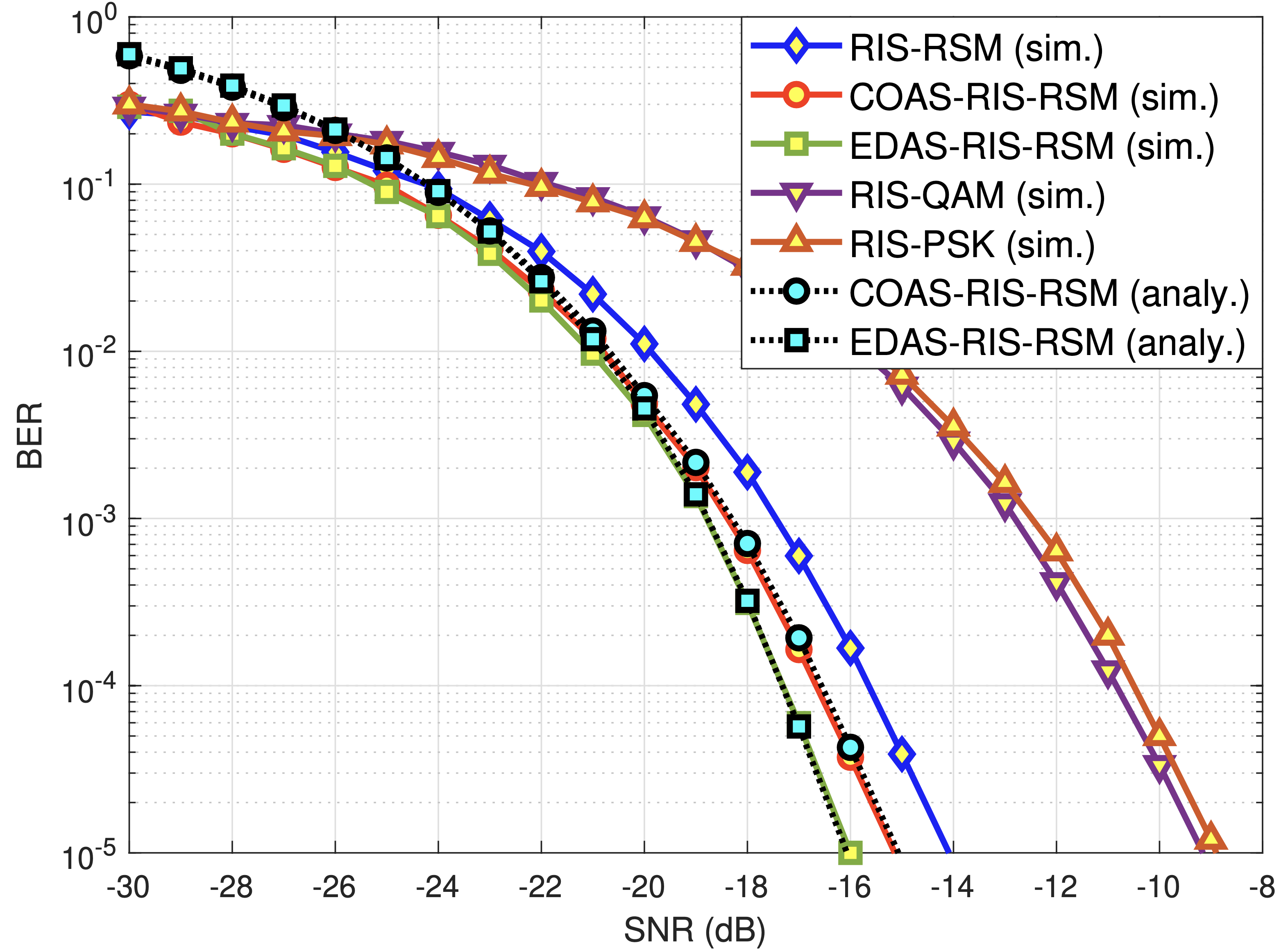}}
	\caption{Performance comparisons of the RIS-PSK, RIS-QAM, RIS-RSM, COAS-RIS-RSM, and EDAS-RIS-RSM systems for $N=32$ while $\eta=3$ bits.}
	\label{AS-RIS-RSM-n3} 
\end{figure}

\begin{figure}[t]
\centering{\includegraphics[width=0.48\textwidth]{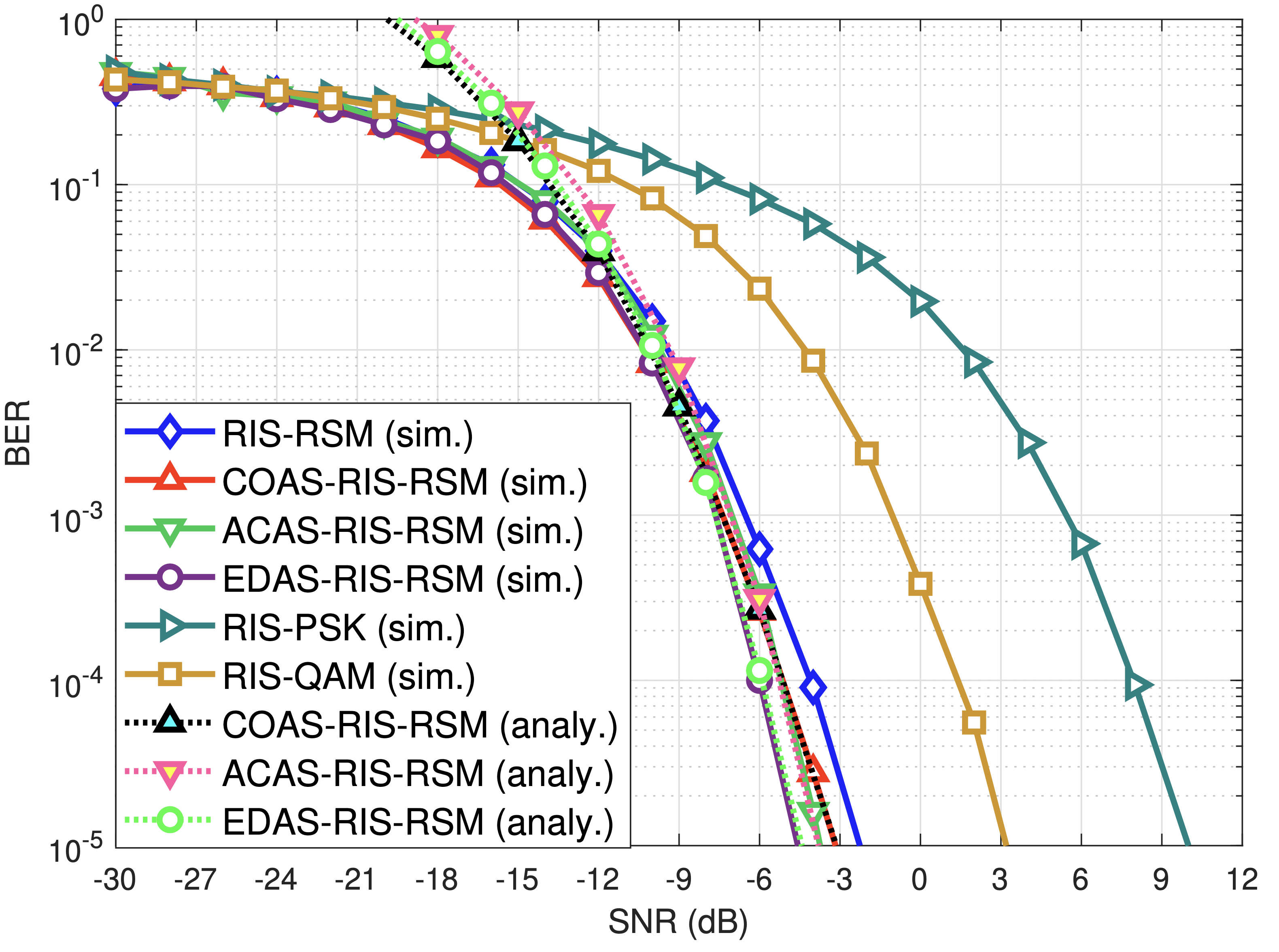}}
	\caption{Performance comparisons of the RIS-PSK, RIS-QAM, RIS-RSM, COAS-RIS-RSM, and EDAS-RIS-RSM systems for $N=16$ while $\eta=5$ bits.}
	\label{AS-RIS-RSM-n5} 
\end{figure}

The BER performance comparisons of the proposed COAS-RIS-RSM and EDAS-RIS-RSM systems compared to the traditional RIS-QAM, RIS-PSK, and RIS-RSM systems are presented for $N=32$ while $\eta=3$ bits in Fig. \ref{AS-RIS-RSM-n3}. The system parameters ($M=4$, $n_R=8$, $n_S=2$) are employed for the proposed COAS-RIS-RSM, ACAS-RIS-RSM, and EDAS-RIS-RSM systems, whereas the RIS-RSM, RIS-QAM, and RIS-PSK systems utilize the system parameters ($M=4$, $n_R=2$), ($M=8$, $n_R=1$), and ($M=8$, $n_R=1$), respectively. Fig. \ref{AS-RIS-RSM-n3} shows that the simulation and analytical results of the proposed COAS-RIS-RSM, ACAS-RIS-RSM, and EDAS-RIS-RSM systems overlap each other. Fig. \ref{AS-RIS-RSM-n3} also shows that the proposed COAS-RIS-RSM and EDAS-RIS-RSM systems exhibit better BER performance than the RIS-RSM, RIS-QAM, and RIS-PSK systems. For $10^{-5}$ BER level, according to RIS-RSM, RIS-QAM, and RIS-PSK systems, the proposed EDAS-RIS-RSM system has SNR gains of $2.01$ dB, $6.94$ dB, and $7.28$ dB, and the proposed COAS-RIS-RSM system has SNR gains of $1.05$ dB, $5.98$ dB, and $6.32$ dB respectively.

The performance of the EDAS-RIS-RSM, ACAS-RIS-RSM, COAS-RIS-RSM, RIS-RSM, RIS-QAM, and RIS-PSK systems for higher spectral efficiency $\eta=5$, similar to Fig. \ref{AS-RIS-RSM-n3}, is being investigated in Fig. \ref{AS-RIS-RSM-n5} for $N=16$. Fig. \ref{AS-RIS-RSM-n3} includes both simulation and analytical results. Fig. \ref{AS-RIS-RSM-n3} shows that the simulation and analytical results are very close to each other, thus confirming each other. While the proposed COAS-RIS-RSM, ACAS-RIS-RSM, and EDAS-RIS-RSM systems employ the system parameters ($M=8$, $n_R=8$, $n_S=4$), the RIS-RSM, RIS-QAM, and RIS-PSK systems use the system parameters ($M=8$, $n_R=4$), ($M=32$, $n_R=1$), and ($M=32$, $n_R=1$), respectively. It can be easily observed in Fig. \ref{AS-RIS-RSM-n5} that the proposed EDAS-RIS-RSM, ACAS-RIS-RSM, and COAS-RIS-RSM systems are capable of achieving data transmission with lower errors than RIS-RSM, RIS-QAM, and RIS-PSK systems. At the $10^{-5}$ BER level, the proposed EDAS-RIS-RSM system exhibits SNR gains of $2.31$ dB, $7.9$ dB, and $14.56$ dB compared to the RIS-RSM, RIS-QAM, and RIS-PSK systems, respectively, while the proposed ACAS-RIS-RSM system demonstrates SNR gains of $1.48$ dB, $7.2$ dB, and $13.73$ dB, respectively, and the proposed COAS-RIS-RSM system provides SNR gains of $0.91$ dB, $6.5$ dB, and $13.16$ dB, respectively.

\begin{figure}[t]
\centering{\includegraphics[width=0.49\textwidth]{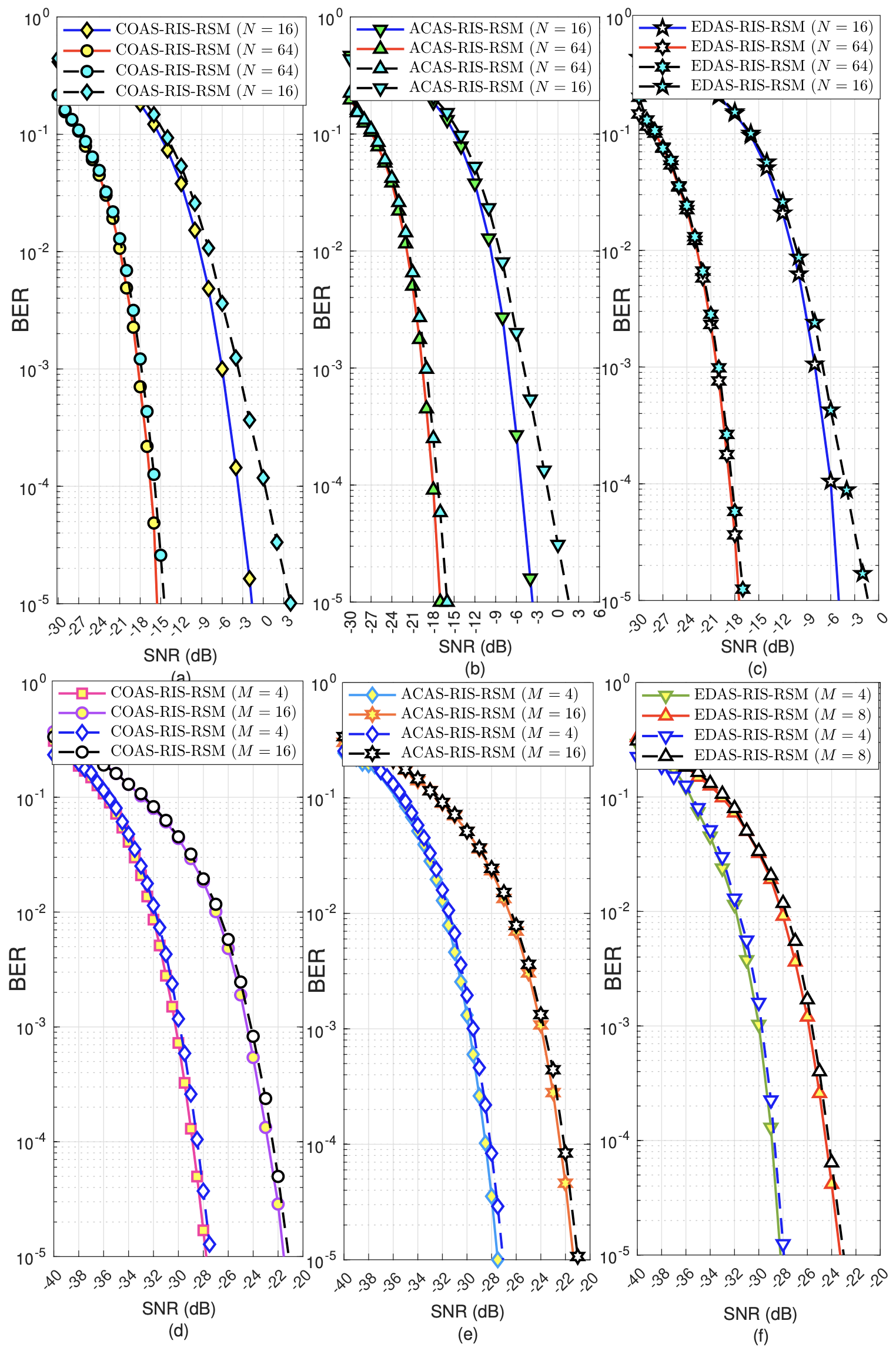}}
	\caption{Performance comparisons of the AS-RIS-RSM systems while $N=16, 64$; (a) COAS-RIS-RSM for $M=16$, $n_R=8$, $n_S=4$, (b) ACAS-RIS-RSM for $M=8$, $n_R=8$, $n_S=4$ and (c) EDAS-RIS-RSM for $M=8$, $n_R=4$, $n_S=2$. Also, performance comparisons of the AS-RIS-RSM systems while $N=128$; (d) COAS-RIS-RSM for $n_R=16$, $n_S=4$, (e) ACAS-RIS-RSM for $n_R=8$, $n_S=4$ and (f) EDAS-RIS-RSM for $n_R=4$, $n_S=2$. Where the continuous line is used for the ML detector and the dashed line for the GD.}
	\label{ALL-AS-RIS-RSM-Ns} 
\end{figure}

The effect of the number of reflecting surface elements, denoted as $N$, on the error performance of the proposed COAS-RIS-RSM, ACAS-RIS-RSM, and EDAS-RIS-RSM systems is examined in Fig. \ref{ALL-AS-RIS-RSM-Ns} (a), (b), and (c) for ML and GD. The system parameters determined for COAS-RIS-RSM, ACAS-RIS-RSM and EDAS-RIS-RSM systems are ($M=16$, $n_R=8$, $n_S=4$), ($M=8$, $n_R=8$, $n_S=4$), and ($M=8$, $n_R=4$, $n_S=2$), respectively. It is shown in Fig. \ref{ALL-AS-RIS-RSM-Ns} (a), (b), and (c) that the ML and GD error performances converge as the number of reflecting surface elements $N$ increases. Since a large number of reflecting surface elements are used to achieve satisfactory error performance improvement in RIS-based wireless communication systems, the GD provides a very close error performance to the ML detector with lower complexity. Fig. \ref{ALL-AS-RIS-RSM-Ns} (a), (b), and (c) shows that the error performance of the proposed systems improves significantly by increasing the number of reflecting surface elements. As can be seen in Fig. \ref{ALL-AS-RIS-RSM-Ns} (a), (b), and (c), when $N$ is quadrupled, the SNR gains of the proposed COAS-RIS-RSM, ACAS-RIS-RSM, and EDAS-RIS-RSM systems increase $13.89$ dB, $13.32$ dB, and $12.47$ dB, respectively for ML detector.

In Fig. \ref{ALL-AS-RIS-RSM-Ns} (d), (e), and (f), the error performance evaluation of the proposed COAS-RIS-RSM, ACAS-RIS-RSM, and EDAS-RIS-RSM systems shows the effect of the modulation order $M$, represented as $M$. The system parameter configurations for the COAS-RIS-RSM, ACAS-RIS-RSM, and EDAS-RIS-RSM systems are selected as ($n_R=16$, $n_S=4$), ($n_R=8$, $n_S=4$), and ($n_R=4$, $n_S=2$), respectively. The error performance of the greedy and ML detectors is quite close to each other since the number of reflecting surface elements is large enough. Also, Fig. \ref{ALL-AS-RIS-RSM-Ns} (d), (e), and (f) distinctly shows the enhancement in the error performance of the proposed systems as the number of $M$ decreases.



\section{Conclusions and Discussions}

This paper presents new wireless communication schemes to improve the performance of wireless SIMO systems through the integration of COAS, ACAS, and EDAS techniques into the RIS-RSM system.  The analytical error performance of the proposed AS-RIS-RSM systems is derived and the results show that the simulation and analytical results overlap each other. The effect of modulation order and number of reflecting surface elements on the error performance of the proposed COAS-RIS-RSM, ACAS-RIS-RSM, and EDAS-RIS-RSM systems is investigated. Also, an optimal ML detector and a sub-optimal GD are presented for estimating the active receive antenna index and symbol at the receiver of the AS-RIS-RSM system. It can be seen that the GD provides lower computational complexity than the ML detector with a certain error performance trade-off. Computational complexity analyses of COAS, ACAS, and EDAS are obtained for the RIS-RSM system. Furthermore, the capacity analysis of the proposed AS-RIS-RSM systems is obtained and compared with RIS and RIS-RSM systems for different system parameters. It is observed that the proposed AS-RIS-RSM systems have higher capacity than RIS and RIS-RSM systems for all cases. Simulation results of the proposed AS-RIS-RSM systems are obtained for Rayleigh fading channels using the $M$-QAM method. As a result, it is observed that the proposed COAS-RIS-RSM, ACAS-RIS-RSM, and EDAS-RIS-RSM systems provide better error performance than RIS-RSM, RIS-QAM, and RIS-PSK systems with a certain complexity cost.


\bibliographystyle{IEEEtran}
\bibliography{IEEEabrv,wileyNJD-AMA}

\end{document}